\documentclass[english,13pt]{article}
\usepackage[T1]{fontenc}
\usepackage[latin1]{inputenc}
\usepackage{geometry}
\usepackage{float}
\usepackage{mathrsfs}
\usepackage{amsmath}
\usepackage{amssymb}
\usepackage{graphicx}
\usepackage{subfigure}

\makeatletter

\usepackage{algorithm,algpseudocode}

\usepackage{amsthm}

\usepackage{mathrsfs}

\usepackage{amsfonts}

\usepackage{epsfig}

\usepackage{bm}

\usepackage{mathrsfs}

\usepackage{enumerate}

\numberwithin{equation}{section}

\@ifundefined{definecolor}{\@ifundefined{definecolor}
 {\@ifundefined{definecolor}
 {\usepackage{color}}{}
}{}
}{}

\usepackage{subfig}\usepackage[all]{xy}

\newtheorem{theorem}{Theorem}[section]
\newtheorem{lem}{Lemma}[section]
\newtheorem{rem}{Remark}[section]

\newcounter{hypA}
\newenvironment{hypA}{\refstepcounter{hypA}\begin{itemize}
  \item[({\bf A\arabic{hypA}})]}{\end{itemize}}
\newcounter{hypB}

\newcounter{hypD}

\usepackage{babel}\date{}

\usepackage{babel}

\makeatother

\usepackage{babel}

\newcommand{\bbE}{\mathbb{E}}

\begin{document}

\begin{center}

{\Large \textbf{An Improved Unbiased Particle Filter}}

\vspace{0.5cm}

BY  AJAY JASRA$^{1}$, MOHAMED MAAMA$^{1}$ \& HERNANDO OMBAO$^{2}$

{\footnotesize $^{1}$Applied Mathematics and Computational Science Program, \footnotesize $^{2}$Statistics Program, Computer, Electrical and Mathematical Sciences and Engineering Division, King Abdullah University of Science and Technology, Thuwal, 23955-6900, KSA.}
{\footnotesize E-Mail:\,} \texttt{\emph{\footnotesize ajay.jasra@kaust.edu.sa, maama.mohamed@gmail.com, hernando.ombao@kaust.edu.sa}}

\end{center}

\begin{abstract}
In this paper we consider the filtering of partially observed multi-dimensional diffusion processes that are observed regularly at discrete times. We assume that, for numerical reasons, one has to time-discretize the diffusion process which typically leads to filtering that is subject to discretization bias. The approach in \cite{ub_pf} establishes that when only having access to the time discretized diffusion it is possible to remove the discretization bias with an estimator of finite variance. We improve on the method in  \cite{ub_pf} by introducing a modified estimator based on the recent work of \cite{anti_mlpf}. We show that this new estimator is unbiased and has finite variance. Moreover, we conjecture and verify in numerical simulations that substantial gains are obtained. That is, for a given mean square error (MSE) and a particular class of multi-dimensional diffusion, the cost to achieve the said MSE falls. 
\end{abstract}

\section{Introduction}

We are given a diffusion process:
\begin{equation}\label{eq:diff_proc}
dX_t = \alpha(X_t)dt + \beta(X_t)dW_t
\end{equation}
where $X_0=x_0\in\mathbb{R}^d$ is given, $\alpha:\mathbb{R}^d\rightarrow\mathbb{R}^d$, $\beta:\mathbb{R}^d\rightarrow\mathbb{R}^{d\times d}$ and $\{W_t\}_{t\geq 0}$ is a
standard $d-$dimensional Brownian motion. This process is unobserved and associated to data that are observed at regular and discrete times. The particular structure of the observations we consider, yields a special case of a state-space or hidden Markov model and the objective is to filter the process at observation times. This class of models has a wide class of applications; see \cite{cappe,delmoral1} for example.

The filtering of diffusion processes requires the application of (stochastic) numerical methods often based on Monte Carlo methods
and moreover by using time-discretization methods such as Euler-Maruyama and the Milstein scheme; see e.g.~\cite{delm,mlpf,ml_rev,anti_mlpf} for some reasons why this is the case. The objective of this article is to obtain (stochastic) estimators of the filter at each observation time, such that one can remove the time-discretization and moreover that this estimator will have finite variance. 

To the best of our knowledge there are two strands of research to obtain unbiased estimates of filters associated to diffusion processes. The first is based upon exact simulation methods (see e.g.~\cite{beskos1,blanchet,fearn}), but such techniques may only apply to a restrictive class of diffusions or be too expensive to implement for filtering. The second class is based upon randomization schemes (indeed so is \cite{blanchet}) that have been developed in \cite{mcl,rhee}; see \cite{vihola} for a nice summary. These latter methods are related to the multilevel Monte Carlo (MLMC) approach \cite{giles,giles1,hein}, so we shall detail these so as to provide a complete picture of the literature. MLMC is a method that is often associated to the approximation of a probability measure that is subject to a `consistent' bias (such as time-discretization). That is, one wishes to calculate an expectation w.r.t.~a pre-specified probability measure, but for often computational reasons, one can only work with an approximation, but as the latter becomes more precise, so does this approximation converge to the expectation of interest. The approach is then to rewrite the expectation w.r.t.~a precise approximation as a collapsing sum of increasingly coarse approximations and then use Monte Carlo methods to further numerically calculate the collapsing sum identity. The key to the methodology is being able to sample from couplings of consecutive (in some sense) approximations and if this latter coupling is appropriately good, the MLMC method can improve upon Monte Carlo, in the sense that the cost to achieve a pre-specified mean square error (MSE) is reduced; we refer to \cite{giles1} for further details. The randomization methods of \cite{mcl,rhee} can be thought of as `randomized MLMC' in that by making the total approximation bias stochastic, one can remove this latter object completely. In some scenarios, these randomizations can have comparable expected costs to MLMC. Most of the original methods in the literature \cite{giles,giles1,ml_anti,mcl,rhee} are based upon exact simulation of the approximated laws, which is typically not possible in the context of filtering.

For filtering of partially observed diffusion processes, several methods based upon particle filters and multilevel particle filters \cite{mlpf,mlpf1,levymlpf,cts_ml} have appeared in the literature. The objective of these ideas is to replicate the success of MLMC when one has to resort to particle filter-type methodology, versus direct sampling.
These methods have been extended to the case of unbiased filtering in the work of \cite{ub_pf} which provides unbiased and finite variance estimators of the filter, recursively in time. In the recent article of \cite{anti_mlpf}, which is based upon \cite{ml_anti}, the authors improved the multilevel particle filter in \cite{mlpf}, in the sense that for 
multi-dimensional diffusion processes with non-constant diffusion coefficients, to achieve a pre-specified MSE, the cost is reduced. This article seeks to combine the method of \cite{ub_pf} with the estimators that are derived in \cite{anti_mlpf}. More specifically, the contributions of this article are:
\begin{enumerate}
\item{Combine the methods of \cite{ub_pf} and \cite{anti_mlpf} to provide unbiased and recursive in time estimates of the filter. }
\item{Under assumptions, we prove that this new estimator is not only unbiased, but can be constructed to have finite variance.}
\item{Under a reasonable conjecture, we show that our new estimator can improve upon that in \cite{ub_pf} for the filtering of multi-dimensional diffusion processes with non-constant diffusion coefficients.}
\item{The claims of 3.~are verified in several numerical examples.}
\end{enumerate}
In terms of 1.~we remark that just as in \cite{ub_pf}, the new estimators are embarrassingly parallel, in that they can be computed at great speed on multiple processors. For 2.~note that just as in \cite{ub_pf} these new estimators have an infinite expected cost. In more details, for 3., let $\epsilon >0$, if one targets an MSE of $\mathcal{O}(\epsilon^2)$, we would have a cost (with high probability - see Section \ref{sec:math} for more discussion on this concept) of $\mathcal{O}(\epsilon^{-2}|\log(\epsilon)|^{2+\delta})$, $\delta>0$, whereas for the method the cost is, with high probability, $\mathcal{O}(\epsilon^{-2.5}|\log(\epsilon)|^{2.5+\delta})$.

This article is structured as follows. In Section \ref{sec:model} we provide details on the model and the time-discretization associated to the process \eqref{eq:diff_proc}. In Section \ref{sec:pf_cpf} we detail existing algorithms needed to develop our method.
Section \ref{sec:ubpf} we present our approach and within this section we also consider our mathematical results. Section \ref{sec:numerics}
we provides numerical results that support our theory.  Our mathematical proofs can be found in the appendix.

\section{Model and Discretization}\label{sec:model}

The model that is considered in this Section is identical to \cite{anti_mlpf}. To that end we use a very similar exposition to that paper below.

\subsection{State-Space Model}

We consider the filtering problem for partially observed diffusions as in \eqref{eq:diff_proc}.  Under assumptions in the appendix this afore-mentioned diffusion has a unique solution and transition 
probability which we denote, over 1 unit time, as $P(x,dx')$.  We set
$$
p(dx_{1:n},y_{1:n}) = \prod_{k=1}^n P(x_{k-1},dx_k)g(x_k,y_k)
$$
where $y_{1:n}=(y_1,\dots,y_n)^{\top}\in\mathsf{Y}^n$ are observations with conditional density $g(x_k,\cdot)$. The filter for $k\in\mathbb{N}$
$$
\pi_k(dx_k) = \frac{\int_{\mathsf{X}^{k-1}} p(dx_{1:k},y_{1:k})}{\int_{\mathsf{X}^{k}} p(dx_{1:k},y_{1:k})}
$$
where $\mathsf{X}=\mathbb{R}^d$. 
Working directly with $P$ is not possible in many cases. We will work with time-discretized filters and then seek to remove the discretization using the approach in \cite{ub_pf}.

\subsection{Time Discretization}

We consider a time discretization at equally spaced times, separated by $\Delta_l=2^{-l}$. Define the $d-$vector, $H:\mathbb{R}^{2d}\times\mathbb{R}^+\rightarrow\mathbb{R}^d$, 
$H_{\Delta}(x,z)=(H_{\Delta,1}(x,z),\dots,H_{\Delta,d}(x,z))^{\top}$ where for $i\in\{1,\dots,d\}$
\begin{eqnarray*}
H_{\Delta,i}(x,z) & = &  \sum_{(j,k)\in\{1,\dots,d\}^2} h_{ijk}(x)(z_jz_k-\Delta) \\
h_{ijk}(x) & = & \frac{1}{2}\sum_{m\in\{1,\dots,d\}} \beta_{mk}(x)\frac{\partial \beta_{ij}(x)}{\partial x_m}.
\end{eqnarray*}
We denote by $\mathcal{N}_d(\mu,\Sigma)$ the $d-$dimensional Gaussian distribution with mean vector $\mu$
and covariance matrix $\Sigma$; if $d=1$ we drop the subscript $d$. $I_d$ is the $d\times d$ identity matrix.
The truncated Milstein scheme, is presented in Algorithm \ref{alg:milstein_sl} and
the antithetic truncated Milstein scheme is Algorithm \ref{alg:milstein}. The method in Algorithm \ref{alg:milstein} is 
an essential ingredient of our subsequent methodology.

We denote the transition kernel induced by Algorithm \ref{alg:milstein_sl} as $P^l(x,dx')$ and for a given $l\in\mathbb{N}_0$ with  we will be concerned with the filter induced by the following joint measure
$$
p^l(dx_{1:n},y_{1:n}) = \prod_{k=1}^n P^l(x_{k-1},dx_k) g(x_k,y_k).
$$
The filter associated to this measure is for $k\in\mathbb{N}$
$$
\pi_k^l(dx_k) = \frac{\int_{\mathsf{X}^{k-1}} p^l(dx_{1:k},y_{1:k})}{\int_{\mathsf{X}^{k}} p^l(dx_{1:k},y_{1:k})}.
$$

\begin{algorithm}
\begin{enumerate}
\item{Input level $l$ and starting point $x_0^l$.}
\item{Generate $Z_k\stackrel{\textrm{i.i.d.}}{\sim}\mathcal{N}_d(0,\Delta_l I_d)$, $k\in\{1,2,\dots,\Delta_l^{-1}\}$.}
\item{Generate level $l$: for $k\in\{0,1,\dots,\Delta_l^{-1}-1\}$ with $X_0^l=x_0^l$
$$
X_{(k+1)\Delta_l}^l = X_{k\Delta_l}^l + \alpha(X_{k\Delta_l}^l)\Delta_l + \beta(X_{k\Delta_l}^l)Z_{k+1} + H_{\Delta_l}(X_{k\Delta_l}^l,Z_{k+1}).
$$
}
\item{Output $X_1^l$.}
\end{enumerate}
\caption{Truncated Milstein Scheme on $[0,1]$.}
\label{alg:milstein_sl}
\end{algorithm}

\begin{algorithm}
\begin{enumerate}
\item{Input level $l$ and starting points $(x_0^l,x_0^{l-1},x_0^{l,a})$.}
\item{Generate $Z_k\stackrel{\textrm{i.i.d.}}{\sim}\mathcal{N}_d(0,\Delta_l I_d)$, $k\in\{1,2,\dots,\Delta_l^{-1}\}$.}
\item{Generate level $l$: for $k\in\{0,1,\dots,\Delta_l^{-1}-1\}$ with $X_0^l=x_0^l$
$$
X_{(k+1)\Delta_l}^l = X_{k\Delta_l}^l + \alpha(X_{k\Delta_l}^l)\Delta_l + \beta(X_{k\Delta_l}^l)Z_{k+1} + H_{\Delta_l}(X_{k\Delta_l}^l,Z_{k+1}).
$$
}
\item{Generate level $l-1$: for $k\in\{0,1,\dots,\Delta_{l-1}^{-1}-1\}$ with $X_0^{l-1}=x_0^{l-1}$
\begin{eqnarray*}
X_{(k+1)\Delta_{l-1}}^{l-1} & = & X_{k\Delta_{l-1}}^{l-1} + \alpha(X_{k\Delta_{l-1}}^{l-1})\Delta_{l-1} + \beta(X_{k\Delta_{l-1}}^{l-1})\{Z_{2(k+1)-1}+Z_{2(k+1)}\} +\\ & & H_{\Delta_{l-1}}(X_{k\Delta_{l-1}}^{l-1},Z_{2(k+1)-1}+Z_{2(k+1)}).
\end{eqnarray*}
}
\item{Generate antithetic level $l$: for $k\in\{0,1,\dots,\Delta_l^{-1}-1\}$ with $X_0^{l,a}=x_0^{l,a}$
$$
X_{(k+1)\Delta_l}^{l,a} = X_{k\Delta_l}^{l,a} + \alpha(X_{k\Delta_l}^{l,a})\Delta_l + \beta(X_{k\Delta_l}^{l,a})Z_{\rho_k} + H_{\Delta_l}(X_{k\Delta_l}^{l,a},Z_{\rho_{k}})
$$
where $\rho_k=k+2\mathbb{I}_{\{0,2,4,\dots\}}(k)$.}
\item{Output $(X_1^l,X_1^{l-1},X_1^{l,a})$.}
\end{enumerate}
\caption{Antithetic Truncated Milstein Scheme on $[0,1]$.}
\label{alg:milstein}
\end{algorithm}

\section{Review of Existing Methods}\label{sec:pf_cpf}

This section follows the exposition that is in \cite{anti_mlpf} and is a requirement to present our new methodology.
Let $\varphi\in\mathcal{B}_b(\mathsf{X})$ with the latter the collection of bounded and measurable real-valued functions: we write $\pi_k^l(\varphi)=\int_{\mathsf{X}}\varphi(x_k)\pi_k^l(dx_k)$
and $\pi_k(\varphi)=\int_{\mathsf{X}}\varphi(x_k)\pi_k(dx_k)$. The objective is to approximate $\pi_k(\varphi)$
using $\pi_k^l(\varphi)$ and we detail how first we can approximate $\pi_k^{\underline{L}}(\varphi)$ for some $\underline{L}\in\mathbb{N}_0$ given and then how to approximate $[\pi_{k}^l-\pi_{k}^{l-1}](\varphi)$.

%

We begin first with approximating $\pi_k^{\underline{L}}(\varphi)$ using the PF as described in Algorithm \ref{alg:pf}.  We note that
$$
\pi_k^{N_{\underline{L}},\underline{L}}(\varphi) := \sum_{i=1}^{N_{\underline{L}}}\frac{g(X_k^{i,\underline{L}},y_k)}{\sum_{j=1}^{N_{l}} g(X_k^{j,\underline{L}},y_k)}\varphi(X_k^{i,\underline{L}})
$$
will converge almost surely to $\pi_k^{\underline{L}}(\varphi)$ where the samples $X_k^{i,\underline{L}}$ are after Step 1.~or 3.~of Algorithm \ref{alg:pf}.

\begin{algorithm}[h]
\begin{enumerate}
\item{Initialization: For $i\in\{1,\dots,N_{\underline{L}}\}$, generate $X_1^{i,\underline{L}}$ independently using Algorithm \ref{alg:milstein_sl} with level $\underline{L}$ and starting point $x_0$. Set $k=1$.}
\item{Resampling: Compute 
\begin{equation}\label{eq:res_weights}
\left(\frac{g(x_k^{1,\underline{L}},y_k)}{\sum_{j=1}^{N_{\underline{L}}} g(x_k^{j,\underline{L}},y_k)},\dots,\frac{g(x_k^{N_{\underline{L}},\underline{L}},y_k)}{\sum_{j=1}^{N_{\underline{L}}} g(x_k^{j,\underline{L}},y_k)}\right).
\end{equation}
For $i\in\{1,\dots,N_{\underline{L}}\}$ generate an index $a_k^i$ using the probability mass function in \eqref{eq:res_weights} and set $\tilde{X}_k^{i,\underline{L}}=X_k^{a_k^i,\underline{L}}$.
Then set $X_k^{1:N_{\underline{L}},\underline{L}}=\tilde{X}_k^{1:N_{\underline{L}},\underline{L}}$.
}
\item{Sampling:  For $i\in\{1,\dots,N_{\underline{L}}\}$, generate $X_{k+1}^{i,\underline{L}}|X_k^{1:N_{\underline{L}},\underline{L}}$ conditionally independently using Algorithm \ref{alg:milstein_sl} with level $\underline{L}$ and starting point $x_k^{i,\underline{L}}$. Set $k=k+1$.}
\end{enumerate}
\caption{Particle Filter at Level $\underline{L}$.}
\label{alg:pf}
\end{algorithm}

Approximating the differences $[\pi_{k}^l-\pi_{k}^{l-1}](\varphi)$ can be peformed using the approach in \cite{anti_mlpf}. That paper gives a new resampling method which is in Algorithm \ref{alg:max_coup}. This resampling method gives rise to the coupled particle filter of \cite{anti_mlpf} which is described in Algorithm \ref{alg:coup_pf}.  We define for $(k,l,N_l,\varphi)\in\mathbb{N}^3\times\mathcal{B}_b(\mathsf{X})$:
\begin{eqnarray}
\pi_k^{N_l,l}(\varphi) & := & \sum_{i=1}^{N_{l}} W_k^{i,l}\varphi(X_k^{i,l}) \label{eq:pi_l_def}\\
\bar{\pi}_k^{N_l,l-1}(\varphi) & := & \sum_{i=1}^{N_{l}} \bar{W}_k^{i,l-1}\varphi(X_k^{i,l-1}) \nonumber\\
\pi_k^{N_l,l,a}(\varphi) & := & \sum_{i=1}^{N_{l}} W_k^{i,l,a}\varphi(X_k^{i,l,a}) \nonumber
\end{eqnarray}
and estimate $[\pi_{k}^l-\pi_{k}^{l-1}](\varphi)$  as
$$
[\pi_{k}^l-\pi_{k}^{l-1}]^{N_l}(\varphi) := \frac{1}{2}\{\pi_k^{N_l,l}(\varphi) + \pi_k^{N_l,l,a}(\varphi)\} - \bar{\pi}_k^{N_l,l-1}(\varphi)
$$
The samples in $(\pi_k^{N_l,l}(\varphi), \pi_k^{N_l,l,a}(\varphi),\bar{\pi}_k^{N_l,l-1}(\varphi))$ are obtained after Step 1.~or 3.~in Algorithm \ref{alg:coup_pf}.


\begin{algorithm}[h]
\begin{enumerate}
\item{Input: $(U_1^{1:N},U_2^{1:N},U_3^{1:N})$ and probabilities $(W_1^{1:N},W_2^{1:N},W_3^{1:N})$.}
\item{For $i\in\{1,\dots,N\}$ generate $U\sim\mathcal{U}_{[0,1]}$ (uniform distribution on $[0,1]$)
\begin{itemize}
\item{If $U<\sum_{i=1}^N \min\{W_1^i,W_2^i,W_3^i\}$ generate $a^i\in\{1,\dots,N\}$ using the probability mass function
$$
\mathbb{P}(i) = \frac{\min\{W_1^i,W_2^i,W_3^i\}}{\sum_{j=1}^N \min\{W_1^j,W_2^j,W_3^j\}}
$$
and set $\tilde{U}_j^i=U_j^{a^i}$, 
$j\in\{1,2,3\}$.}
\item{Otherwise generate $(a_1^i,a_2^i,a_3^i)\in\{1,\dots,N\}^3$ using any coupling of the probability mass functions:
$$
\mathbb{P}_j(i) = \frac{W_j^i-\min\{W_1^i,W_2^i,W_3^i\}}{\sum_{k=1}^N[W_j^k-\min\{W_1^k,W_2^k,W_3^k\}]}
$$
and set $\tilde{U}_j^i=U_j^{a_j^i}$,  $j\in\{1,2,3\}$.}
\end{itemize}
}
\item{Set: $U_j^i=\tilde{U}_j^{i}$, $(i,j)\in\{1,\dots,N\}\times\{1,2,3\}$.}
\item{Output: $(U_1^{1:N},U_2^{1:N},U_3^{1:N})$.}
\end{enumerate}
\caption{Maximal Coupling-Type Resampling}
\label{alg:max_coup}
\end{algorithm}

\begin{algorithm}[h]
\begin{enumerate}
\item{Initialization: For $i\in\{1,\dots,N_{\underline{l}}\}$, generate $U_1^{i,l} = (X_1^{i,l},\bar{X}_1^{i,l-1},X_1^{i,l,a})$ independently using Algorithm \ref{alg:milstein} with level $l$ and starting points $(x_0^{l},\bar{x}_0^{l-1},x_0^{l,a})$. Set $k=1$.}
\item{Coupled Resampling: Compute 
\begin{eqnarray*}
W_k^{1:N_l,l} & = &  \left(\frac{g(x_k^{1,l},y_k)}{\sum_{j=1}^{N_{l}} g(x_k^{j,l},y_k)},\dots,\frac{g(x_k^{N_l,l},y_k)}{\sum_{j=1}^{N_{l}} g(x_k^{j,l},y_k)}\right) \\
\bar{W}_k^{1:N_l,l-1} & = &  \left(\frac{g(\bar{x}_k^{1,l-1},y_k)}{\sum_{j=1}^{N_{l}} g(\bar{x}_k^{j,l-1},y_k)},\dots,\frac{g(\bar{x}_k^{N_l,l-1},y_k)}{\sum_{j=1}^{N_{l}} g(\bar{x}_k^{j,l-1},y_k)}\right) \\
W_k^{1:N_l,l,a} & = & \left(\frac{g(x_k^{1,l,a},y_k)}{\sum_{j=1}^{N_{l}} g(x_k^{j,l,a},y_k)},\dots,\frac{g(x_k^{N_l,l,a},y_k)}{\sum_{j=1}^{N_{l}} g(x_k^{j,l,a},y_k)}\right).
\end{eqnarray*}
Then using
$(X_k^{1:N_l,l},\bar{X}_k^{1:N_l,l-1}, X_k^{1:N_l,l,a})$ and $(W_k^{1:N_l,l},\bar{W}_k^{1:N_l,l-1},W_k^{1:N_l,l,a})$ call Algorithm \ref{alg:max_coup}.
}
\item{Coupled Sampling: For $i\in\{1,\dots,N_{l}\}$, generate $U_{k+1}^{i,l}|U_k^{1:N_{l},l}$ conditionally independently using Algorithm \ref{alg:milstein} with level $l$ and starting point $(x_k^{i,l},\bar{x}_k^{i,l-1},x_k^{i,l,a})$. Set $k=k+1$.}
\end{enumerate}
\caption{Coupled Particle Filter of \cite{anti_mlpf} for $l\in\mathbb{N}$ given.}
\label{alg:coup_pf}
\end{algorithm}

\section{Unbiased Particle Filter}\label{sec:ubpf}

\subsection{Overarching Approach}

We are in the context where for any $(k,\varphi)\in\mathbb{N}\times\mathcal{B}_b(\mathsf{X})$ (see e.g.~\cite{anti_mlpf})
\begin{equation}\label{eq:weak_error_gen}
\lim_{l\rightarrow\infty}\pi_k^l(\varphi)=\pi_k(\varphi).
\end{equation}

We will develop a Monte Carlo method that can deliver unbiased and finite variance estimates of $\pi_k(\varphi)$, using randomization approaches (e.g.~\cite{rhee}, see also \cite[Theorem 3]{vihola}). 
It is assumed that one can produce
a sequence of independent random variables $(\Xi_l)_{l\in\mathbb{N}_{\underline{L}}}$, $\mathbb{N}_{\underline{L}}:=\{\underline{L},\underline{L}+1,\dots\}$  such that 
for any $k\in\mathbb{N}$
(subscript $k$ is suppressed from $\Xi_l$ for readability)
\begin{equation}\label{eq:xi_cond}
\mathbb{E}[\Xi_l] = 
\pi_k^l(\varphi) - \pi_k^{l-1}(\varphi)  = [\pi_k^l-\pi_k^{l-1}](\varphi)
\end{equation}
where  $\pi_k^{\underline{L}-1}(\varphi):=0$.
Let $\mathbb{P}_L(l)$ be a positive probability mass function on $\mathbb{N}_{\underline{L}}$. 
If $L$ is simulated from $\mathbb{P}_L$ and one considers the estimate 
\begin{equation}\label{eq:eta_singleterm}
\widehat{\pi_k(\varphi)}_1 = 
\frac{\Xi_L}{\mathbb{P}_L(L)} \, .
\end{equation}
By \cite[Theorem 3]{vihola}, $\widehat{\pi_k(\varphi)}_1$ is an unbiased and finite variance estimator of $\eta(\varphi)$ if
\begin{equation}\label{eq:ub_fv_cond}
\sum_{l\in\mathbb{N}_{\underline{L}}}\frac{1}{\mathbb{P}_L(l)}\mathbb{E}[\Xi_l^2] < +\infty \, .
\end{equation}
Notice 
that once we can obtain \eqref{eq:eta_singleterm}, 
we can construct i.i.d.~samples in parallel by sampling $L_i$ independently from $\mathbb{P}_L$ for 
$i=1,\dots, M$. 
From these samples we are able to construct an unbiased estimator with 
$\mathcal{O}(M^{-1})$ mean squared error 
(assuming condition \eqref{eq:ub_fv_cond} holds) as follows
$$
\widehat{\pi_k(\varphi)} = \frac1M \sum_{i=1}^M \frac{\Xi_{L_i}}{\mathbb{P}_L(L_{i})} \, .
$$

The objective is to deliver an algorithm which can provide unbiased estimates of $\pi_k^l(\varphi)$, $(\Xi_l)_{l\in\mathbb{N}_{\underline{L}}}$ and $\mathbb{P}_L$ such that \eqref{eq:xi_cond} and \eqref{eq:ub_fv_cond} are satisfied. This was solved in \cite{ub_pf} and we show how this can be improved using the methodology in Section \ref{sec:pf_cpf}.

\subsection{Strategy of \cite{ub_pf}}

We now describe how to obtain $(\Xi_l)_{l\in\mathbb{N}_{\underline{L}}}$ via unbiased estimates of $\bar{\pi}_k^{\underline{L}}(\varphi)$ and of $[\bar{\pi}_k^l-\bar{\pi}_k^{l-1}](\varphi)$. The approach of \cite{ub_pf}
uses the biased estimates $\bar{\pi}_k^{\underline{L}}(\varphi)$ and of $[\bar{\pi}_k^l-\bar{\pi}_k^{l-1}](\varphi)$
which are from the particle filter and coupled particle filter respectively. The latter coupled particle filter used in \cite{ub_pf} is the one in \cite{mlpf}, whereas, we use the one in \cite{anti_mlpf} which was described in Section \ref{sec:pf_cpf}.

The approach in \cite{ub_pf} is as follows.
Let $(N_p)_{p\in\mathbb{N}_0}$, $N_p\in\mathbb{N}$  be an increasing sequence of positive integers, with $\lim_{p\rightarrow\infty} N_p=\infty$. 
If, almost surely, we have some Monte Carlo estimators based on $N_p$ samples such that
$$
\lim_{p\rightarrow\infty}\widehat{\pi}_k^{N_p,\underline{L}}(\varphi) = \pi_k^{\underline{L}}(\varphi) \, ,
\quad {\rm and} \quad  
\lim_{p\rightarrow\infty}\widehat{[\pi_k^{l}-\pi_k^{l-1}]}^{N_p}(\varphi)=[\pi_k^{l}-\pi_k^{l-1}](\varphi).
$$
Below we will define estimators of $\pi_k^{\underline{L}}(\varphi)$ ($\widehat{\pi}_k^{N_p,\underline{L}}(\varphi)$) and $[\pi_k^{l}-\pi_k^{l-1}](\varphi)$ ($\widehat{[\pi_k^{l}-\pi_k^{l-1}]}^{N_p}(\varphi)$)
using the methods in Section \ref{sec:pf_cpf}. If $N_p=0$ then we take the, yet to be deifned, estimators as identically zero.


Suppose that one has a positive probability mass function $\mathbb{P}_P(p)$ on $p\in\mathbb{N}_0$.
Define
$$
\Xi_{l,p} :=  \left\{\begin{array}{ll}
 \frac{1}{\mathbb{P}_P(p)}[
\widehat{\pi}_k^{N_p,\underline{L}}(\varphi) - \widehat{\pi}_k^{N_{p-1},\underline{L}}(\varphi)
 & \textrm{if}~l=\underline{L} \\
\frac{1}{\mathbb{P}_P(p)}\Big(
\widehat{[\pi_k^{l}-\pi_k^{l-1}]}^{N_p}(\varphi) - 
\widehat{[\pi_k^{l}-\pi_k^{l-1}]}^{N_{p-1}}(\varphi)
\Big)& \textrm{otherwise} \, 
\end{array}\right .
$$
and $N_{-1}=0$.
Now set $\Xi_l =  \Xi_{l,P}$, 
where $P$ is sampled according to $\mathbb{P}_P(p)$.
Again using \cite[Theorem 3]{vihola}, we have that
$$
\mathbb{E}[\Xi_l] =  \left\{\begin{array}{ll}
\pi_k^{\underline{L}}(\varphi) & \textrm{if}~l=\underline{L} \\
\pi_k^{l}(\varphi)-\pi_k^{l-1}(\varphi) & \textrm{otherwise} \, .
\end{array}\right.
$$
For each $l\in\mathbb{N}_0$, $\Xi_l$ has finite variance provided one has
\begin{equation}\label{eq:xipl_cond}
\bbE [\Xi_l^2] = 
\sum_{p\in\mathbb{N}_0} \mathbb{P}_P(p) \bbE [\Xi_{l,p}^2 ] <+\infty \, .
\end{equation}
Now we describe how to  compute $\widehat{\pi}_k^{N_p,\underline{L}}(\varphi)$ and $\widehat{[\pi_k^{l}-\pi_k^{l-1}]}^{N_p}(\varphi)$ using the ideas in \cite{ub_pf}.

\subsection{Additional Estimators}

Throughout the section $k\in\mathbb{N}$ is given and fixed.  Below, when we call Algorithms \ref{alg:pf} and
\ref{alg:coup_pf} we mean to run it up-to the specified time $k$.
To obtain $\widehat{\pi}_k^{N_0,\underline{L}}(\varphi)$ with $N_0$ samples, we run the PF as 
in Algorithm \ref{alg:pf}. Then to compute $\widehat{\pi}_k^{N_1,\underline{L}}(\varphi)$ we run a PF independently of the first PF with $N_1-N_0$ samples and so on, for any $p\geq 2$ (i.e.~with $N_2-N_1,\dots, N_{p}-N_{p-1}$ samples).
Set
\begin{eqnarray*}
\eta_{k}^{N_{0:p},\underline{L}}(\varphi) & := & \sum_{q=0}^p \Big(\frac{N_q-N_{q-1}}{N_p}\Big)\eta_{k}^{N_q-N_{q-1},\underline{L}}(\varphi) \, ,\\
\eta_{k}^{N_q-N_{q-1},\underline{L}}(\varphi) & := & \frac{1}{N_{q}-N_{q-1}}\sum_{i=N_{q-1}+1}^{N_q}\varphi(x_k^{i,\underline{L}}) \, .
\end{eqnarray*}
Here $x_k^{1,\underline{L}},\dots,x_k^{N_{0},\underline{L}}$ are generated from the first PF, independently $x_k^{N_0+1,\underline{L}},\dots,x_k^{N_{1},\underline{L}}$ from the second and so on.
The procedure for sampling, in order to compute \eqref{eq:eta_0_app} below is in Algorithm \ref{alg:pf_est}.
The approximation of $\widehat{\pi}_k^{N_0,\underline{L}}(\varphi)$ is finally 
\begin{equation}\label{eq:eta_0_app}
\widehat{\pi}_k^{N_0,\underline{L}}(\varphi)
 = 
\frac{\eta_{k}^{N_{0:p},\underline{L}}(g_k\varphi)}{\eta_{k}^{N_{0:p},\underline{L}}(g_k)} 
\end{equation}
where we use the notation $g_k$ for the likelihood of $y_k$ given $x_k$ and suppress $y_k$ from the notation.

\begin{algorithm}
\begin{enumerate}
\item{Initialization: Run Algorithm \ref{alg:pf} with $N_0$ samples. Set $q=1$. 
If $p=0$ stop; otherwise go to 2.}
\item{Iteration: Independently of all other samples, run Algorithm \ref{alg:pf} with 
$N_q-N_{q-1}$ samples. Set $q=q+1$. 
If $q=p+1$ stop; otherwise go to the start of 2.}
\end{enumerate}
\caption{Approach for sampling to compute \eqref{eq:eta_0_app}, for $p\in\mathbb{N}_0$ given.}
\label{alg:pf_est}
\end{algorithm}

Let $l\in\mathbb{N}$ be given.  To obtain $\widehat{[\pi_k^{l}-\pi_k^{l-1}]}^{N_0}(\varphi)$
 we run the CPF in Algorithm \ref{alg:coup_pf} with $N_0$ samples. To form the approximation
with $N_1$ samples, we run a CPF independently of the first CPF with $N_1-N_0$ samples and so on, for any $p\geq 2$ (i.e.~with $N_2-N_1,\dots, N_{p}-N_{p-1}$ samples).
For $s\in\{l,l-1\}$ and any $\varphi\in\mathcal{B}_b(\mathsf{X})$ we define 
\begin{eqnarray*}
\eta_k^{N_{0:p},s}(\varphi) & := & \sum_{q=0}^p\Big(\frac{N_q-N_{q-1}}{N_p}\Big)\eta_k^{N_q-N_{q-1},s}(\varphi) \, , \\
\eta_k^{N_q-N_{q-1},s}(\varphi) & := & \frac{1}{N_q-N_{q-1}}\sum_{i=N_{q-1}+1}^{N_q}\varphi(x_k^{i,s}) \, .
\end{eqnarray*}
The extension of the notation for the antithetic samples is clear (e.g.~$\eta_k^{N_q-N_{q-1},l,a}(\varphi)$) etc.
Then we set 
\begin{equation}\label{eq:eta_l_app}
\widehat{[\pi_k^{l}-\pi_k^{l-1}]}^{N_p}(\varphi) = 
\frac{\tfrac{1}{2}\eta_n^{N_{0:p},l}(g_k\varphi)}{\eta_k^{N_{0:p},l}(g_k)}
+ \frac{\tfrac{1}{2}\eta_n^{N_{0:p},l,a}(g_k\varphi)}{\eta_k^{N_{0:p},l,a}(g_k)}
 - \frac{\eta_k^{N_{0:p},l-1}(g_k\varphi)}{\eta_k^{N_{0:p},l-1}(g_k)} \, .
\end{equation}
The procedure for sampling, in order to compute \eqref{eq:eta_l_app} is summarized in Algorithm \ref{alg:cpf_est}.
It is this estimator, based on that derived in \cite{anti_mlpf}, which differs from that of \cite{ub_pf}.

\begin{algorithm}[h!]
\begin{enumerate}
\item{Initialization: Run Algorithm \ref{alg:coup_pf} with $N_0$ samples. Set $q=1$. If $p=0$ stop, otherwise go to 2.}
\item{Iteration: Independently of all other samples, run Algorithm \ref{alg:coup_pf} with $N_q-N_{q-1}$ samples. Set $q=q+1$. If $q=p+1$ stop; otherwise go to the start of 2.}
\end{enumerate}
\caption{Approach for sampling to compute \eqref{eq:eta_l_app}, for $(p,l)\in\mathbb{N}_0\times\mathbb{N}$ given.}
\label{alg:cpf_est}
\end{algorithm}

\subsection{An Improved Unbiased Particle Filter}

To summarize our method is in Algorithm \ref{alg:main_method}. This is the same as \cite{ub_pf} with the exception of the changes outlined above. 

\begin{algorithm}[h!]
\begin{enumerate}
\item{For $i\in\{1,\dots,M\}$ sample $L_i\in\mathbb{N}_{\underline{L}}$ according to $\mathbb{P}_L$ and $P_i\in\mathbb{N}_0$ according to $\mathbb{P}_P$. 
Denote the realizations as $(l_i, p_i)$.}
\item{If $l_i=\underline{L}$ compute 
$$
\Xi_{l_i,p_i} =  \frac{1}{\mathbb{P}_P(p_i)}\{
\widehat{\pi}_k^{N_{p_i},\underline{L}}(\varphi) -
\widehat{\pi}_k^{N_{p_i-1},\underline{L}}(\varphi)\}
$$
where $\widehat{\pi}_k^{N_{p},\underline{L}}(\varphi)$ is as \eqref{eq:eta_0_app} (see Algorithm \ref{alg:pf_est}).}
\item{Otherwise, compute 
$$
\Xi_{l_i,p_i} =  \frac{1}{\mathbb{P}_P(p_i)}
\Big(\widehat{[\pi_k^{l_i}-\pi_k^{l_{i}-1}]}^{N_{p_i}}(\varphi) -
\widehat{[\pi_k^{l_i}-\pi_k^{l_{i}-1}]}^{N_{p_{i}-1}}(\varphi)
\Big)
$$
where $\widehat{[\pi_k^{l}-\pi_k^{l-1}]}^{N_p}(\varphi)$ is as \eqref{eq:eta_l_app} (see Algorithm \ref{alg:cpf_est}).}
\item{Return the estimate:
\begin{equation}\label{eq:ub_pf_est}
\widehat{\pi_k(\varphi)} = \frac{1}{M}\sum_{i=1}^M \frac{1}{\mathbb{P}_L(l_i)}\Xi_{l_i,p_i}.
\end{equation}
}
\end{enumerate}
\caption{Algorithm for Unbiased Estimation of $\bar{\eta}_n(\varphi)$.}
\label{alg:main_method}
\end{algorithm}

\subsection{Mathematical Results}\label{sec:math}

We now present our main mathematical result. Below, we denote by $\mathcal{C}_b^2(\mathsf{X},\mathbb{R})$ the collection of functions that  are twice continuously differentiable from $\mathsf{X}$ to $\mathbb{R}$ with bounded derivatives of all order 1 and 2.

\begin{theorem}
Assume (A\ref{ass:diff1}-\ref{ass:g}). 
Then for any $(k,\varphi)\in\mathbb{N}\times\mathcal{B}_b(\mathsf{X})\cap\mathcal{C}_b^2(\mathsf{X},\mathbb{R})$
there exist choices of $\mathbb{P}_L$, $\mathbb{P}_p$ and increasing sequence $(N_p)_{p\in\mathbb{N}_0}$
such that \eqref{eq:ub_pf_est} is an unbiased and finite variance estimator
of $\pi_k(\varphi)$.
\end{theorem}

\begin{proof}
The proof is essentially identical to that of \cite[Theorem 2]{ub_pf}, except we use Lemmata \ref{lem:single} \& \ref{lem:coup} in place of \cite[Proposition 2 \& Theorem 1]{ub_pf}. Hence the proof is omitted.
\end{proof}

The implications of this result are not as clear-cut as for \cite[Theorem 2]{ub_pf}. In the afore-mentioned result the proof reveals sensible choices of $\mathbb{P}_L$, $\mathbb{P}_p$ and $(N_p)_{p\in\mathbb{N}_0}$. However, the bounds from \cite{anti_mlpf} are perhaps cruder than those of \cite{mlpf} which drives the result  
\cite[Theorem 2]{ub_pf}. Although the bounds of \cite{anti_mlpf} are very useful for multilevel calculations, as our estimator decouples level and the number of samples used for the coupled particle filter, the bounds are less effective for determining $\mathbb{P}_L$, $\mathbb{P}_p$ and $(N_p)_{p\in\mathbb{N}_0}$. In general, as we will (indirectly) show in simulations we believe that
\begin{equation}\label{eq:exp_var}
\mathbb{E}\left[\left(
\widehat{[\pi_k^{l}-\pi_k^{l-1}]}^{N_p}(\varphi) - [\pi_k^{l}-\pi_k^{l-1}](\varphi)
\right)^2\right] = \mathcal{O}\left(\frac{\Delta_l}{N_p}\left\{1+\frac{p^2}{N_p}\right\}\right).
\end{equation}
In such a case, one can follow exactly the discussion of \cite[Section 3.2]{ub_pf}, to choose 
$\mathbb{P}_L(l)\propto\Delta_l^{-1}(l+1)\log_2(l+2)^2$, $\mathbb{P}_P(p)\propto N_p^{-1}(p+1)\log_2(p+1)^2$ and $N_p=2^p$. Then our estimator is unbiased and of finite variance. Moreover, for $\epsilon>0$ given, setting $M=\mathcal{O}(\epsilon^{-2})$, one can show that the cost (with high-probability - see \cite[Section 4, Column 2]{rhee}) is $\mathcal{O}(\epsilon^{-2}|\log(\epsilon)|^{2+\delta})$. Now, if one consider the case of \eqref{eq:diff_proc} with $d>1$ and $\beta$ non-constant, the method in \cite{ub_pf} would
cost (with high-probability), based upon \cite[Proposition 5]{rhee},  $\mathcal{O}(\epsilon^{-2.5}|\log(\epsilon)|^{2.5+\delta})$. Both methods achieve a mean square error (MSE) of $\mathcal{O}(\epsilon^2)$.
 
 In the scenario where one truncates the discretization level $l$ to $\{\underline{L},\dots,L_{\textrm{max}}\}$ say and that one truncates $p$ to $\{0,\dots,P_{\textrm{max}}\}$ then one has biased but finite (in expectation) cost. The method introduced here would (under the choices of $\mathbb{P}_L$, $\mathbb{P}_p$ and $(N_p)_{p\in\mathbb{N}_0}$given in \cite[Section 4.3]{ub_pf}) have an expected cost of $\mathcal{O}(\epsilon^{-2}|\log(\epsilon)|^{4})$ and (for multi-dimensional SDEs with non-constant diffusion coefficients) the method of \cite{ub_pf} would have an expected cost  $\mathcal{O}(\epsilon^{-2.5}|\log(\epsilon)|^{3})$. These costs would achieve an MSE of $\mathcal{O}(\epsilon^2)$.
 
The discussion above, of course, assumes that \eqref{eq:exp_var} holds, but we believe it is a sensible conjecture. In addition, for multi-dimensional SDEs with non-constant diffusion coefficients, we have argued that the method introduced here improves upon that in \cite{ub_pf}, in the sense that the cost to achieve a given MSE falls. We shall now establish that this holds in several practical examples.

\section{Numerical Results}\label{sec:numerics}

%
\subsection{Models}

We consider three different models for our numerical experiments.

\subsubsection{Model 1: Geometric Brownian motion (GBM) process} 
 Our first model we use is :
$$
dX_t  = \mu X_t dt + \sigma X_t dW_t, \quad X_0=x_0.
$$
We set $Y_n|X_{n}=x \sim \mathcal{N}(\log(x),\tau^2)$ where $\tau^2=0.02$
and $\mathcal{N}(x,\tau^2)$ is the Gaussian distribution with mean $x$ and variance $\tau^2$. We choose $x_0 = 1$, $\mu = 0.02$, and $\sigma = 0.2$.

\subsubsection{Model 2: Clark-Cameron SDE}
 Our second model we consider in this paper is the Clark-Cameron SDE model (e.g.~\cite{ml_anti}) with initial conditions $x_0=(0,0)^{\top}$
\begin{eqnarray*}
dX_{1,t}  & = &  dW_{1,t}    \\ 
dX_{2,t}  & = & X_{1,t}  dW_{2,t}
\end{eqnarray*}
where $X_{j,t}$ denotes the $j^{th}$ dimension of $X_t$, $j\in\{1,\dots,d\}$.
In addition,  $Y_n|X_n=x\sim \mathcal{N}( \tfrac{X_{1} + X_{2}}{2},\tau^2)$ where $\tau^2=0.1$

\subsubsection{Model 3: Multi-dimensional SDEs with a nonlinear diffusion term (NLMs)} 

For our last model we use  the following multi-dimensional SDEs, with $x_0=(0,0)^{\top}$
 \begin{eqnarray*}
dX_{1,t}  & = & \theta_{1} (\mu_{1} - X_{1,t}) dt + \frac{\sigma_1}{\sqrt{1 + X_{1,t}^{2} }}  dW_{1,t}    \\ 
dX_{2,t}  & =  & \theta_{2} (\mu_{2} - X_{1,t}) dt +   \frac{\sigma_2}{\sqrt{1 + X_{1,t}^{2}}}   dW_{2,t}.
\end{eqnarray*}
We set 
$Y_n|X_n=x\sim \mathcal{L}\left(\tfrac{X_{1} + X_{2}}{2},s\right)$ where $\mathcal{L}(m,s)$ is the Laplace distribution with location $m$ and scale $s$. The values of the parameters that we choose are $(\theta_{1}, \theta_{2}) = (1,1)$, $(\mu_{1}, \mu_{2}) = (0,0)$, $(\sigma_{1}, \sigma_{2}) = (1,1)$ and $s= \sqrt{0.1}$.

\subsection{Simulation Settings}

For our numerical experiments, the unbiased estimator in our algorithm will be compared with Antithetic Multilevel Particle Filters (AMLPF) as in \cite{anti_mlpf} and the unbiased estimator of \cite{ub_pf}.
For a given value $\epsilon$, to target an MSE of $\mathcal{O}(\epsilon^2)$ and for controlling the expected cost, we truncate the distribution on $P$ and $L$ for our unbiased estimator. We then set for a truncation value $L_{\textrm{max}}$ given, $P_{\textrm{max}} = L_{\textrm{max}}$ and constrain the support of $l$ and $p$ to $\left\{ 0,1,2,...,L_{\textrm{max}}  \right\}^2$. We choose the following settings 
$$ 
\mathbb{P}_{P}(p) \propto 2^{-p}, \quad \mathbb{P}_{L}(l) \propto 2^{-\tau l}, \quad N_{p} = 2^{p} N_0 , \quad N_0  \propto P_{\textrm{max}}^{2} 2^{2 P_{\textrm{max}}}, \quad \Delta_l = 2^{-l}, \quad M = \mathcal{O}(\epsilon^{-2})
$$
and $\tau\in\{\tfrac{1}{2},1\}$, depending on whether we use the method in \cite{ub_pf} ($\tau=\tfrac{1}{2}$) or the method of this paper ($\tau=1$).
We note that the unbiased estimator \eqref{eq:ub_pf_est} is the average of $M$ i.i.d.~realizations 
and the AMLPF estimator is given by a telescoping sum (see \cite{anti_mlpf}) of particle and coupled particle filters.

The ground truth of the first model (GBM process) is computed by a Kalman filter and For the two other models, results are generated from a high-resolution simulation of particle filter to approximate the ground truth. Resampling is done adaptively and each simulation is repeated $100$ times.

\subsection{Simulation Results}

\begin{figure}[h!]
\centering
\subfigure{\includegraphics[width=12cm,height=5.5cm]{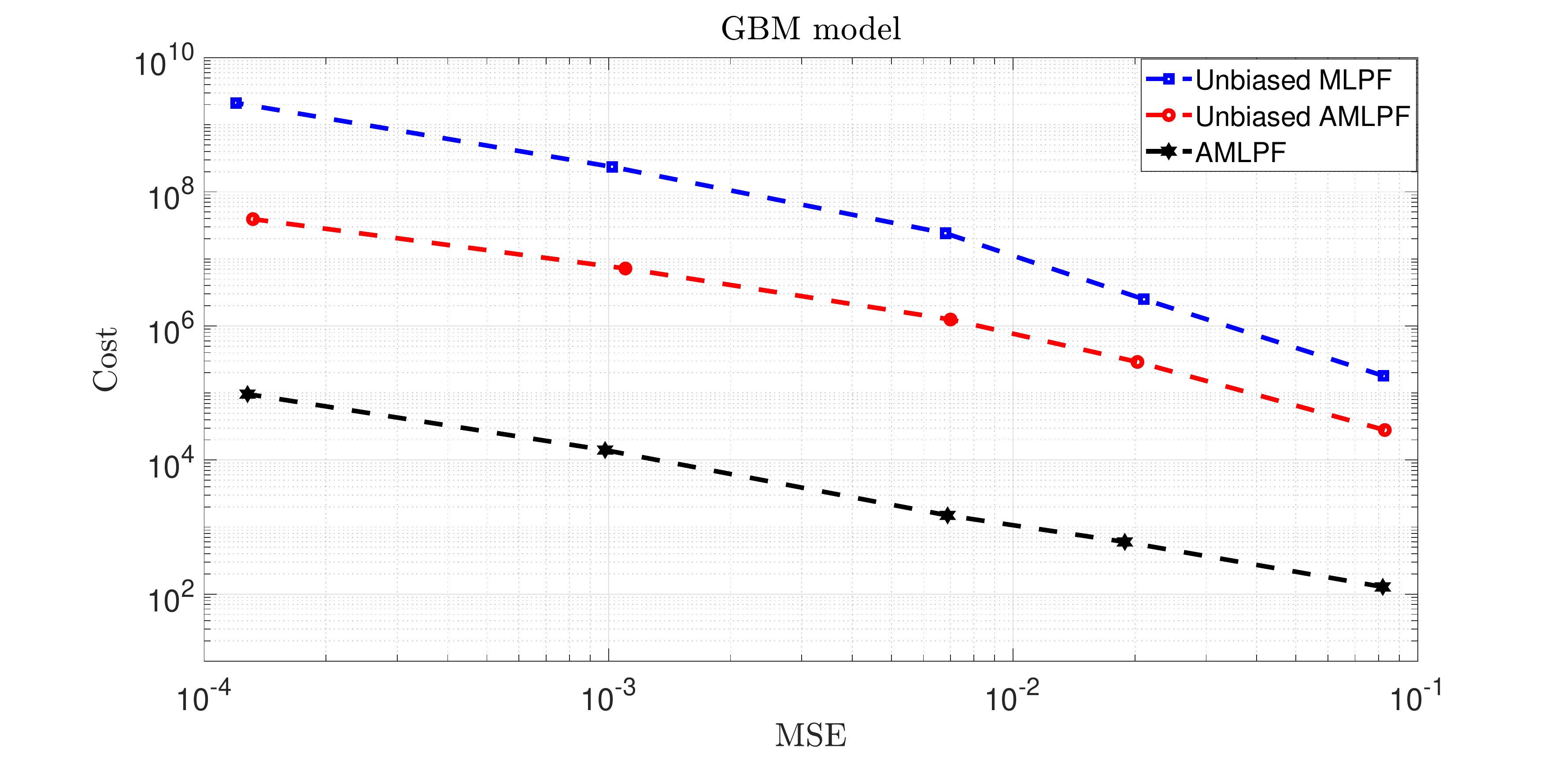}}
\subfigure{\includegraphics[width=12cm,height=5.5cm]{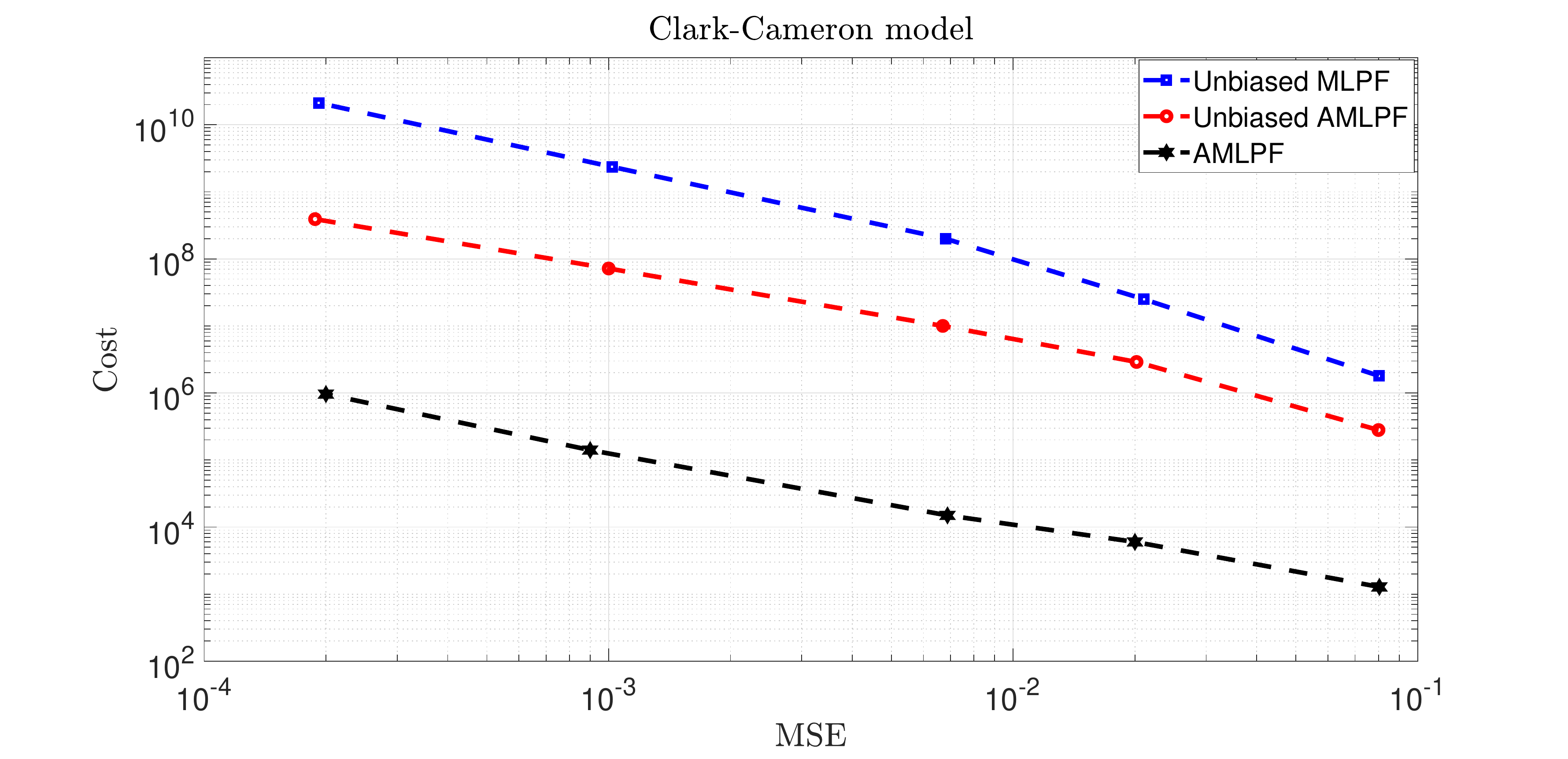}}
\subfigure{\includegraphics[width=12cm,height=5.5cm]{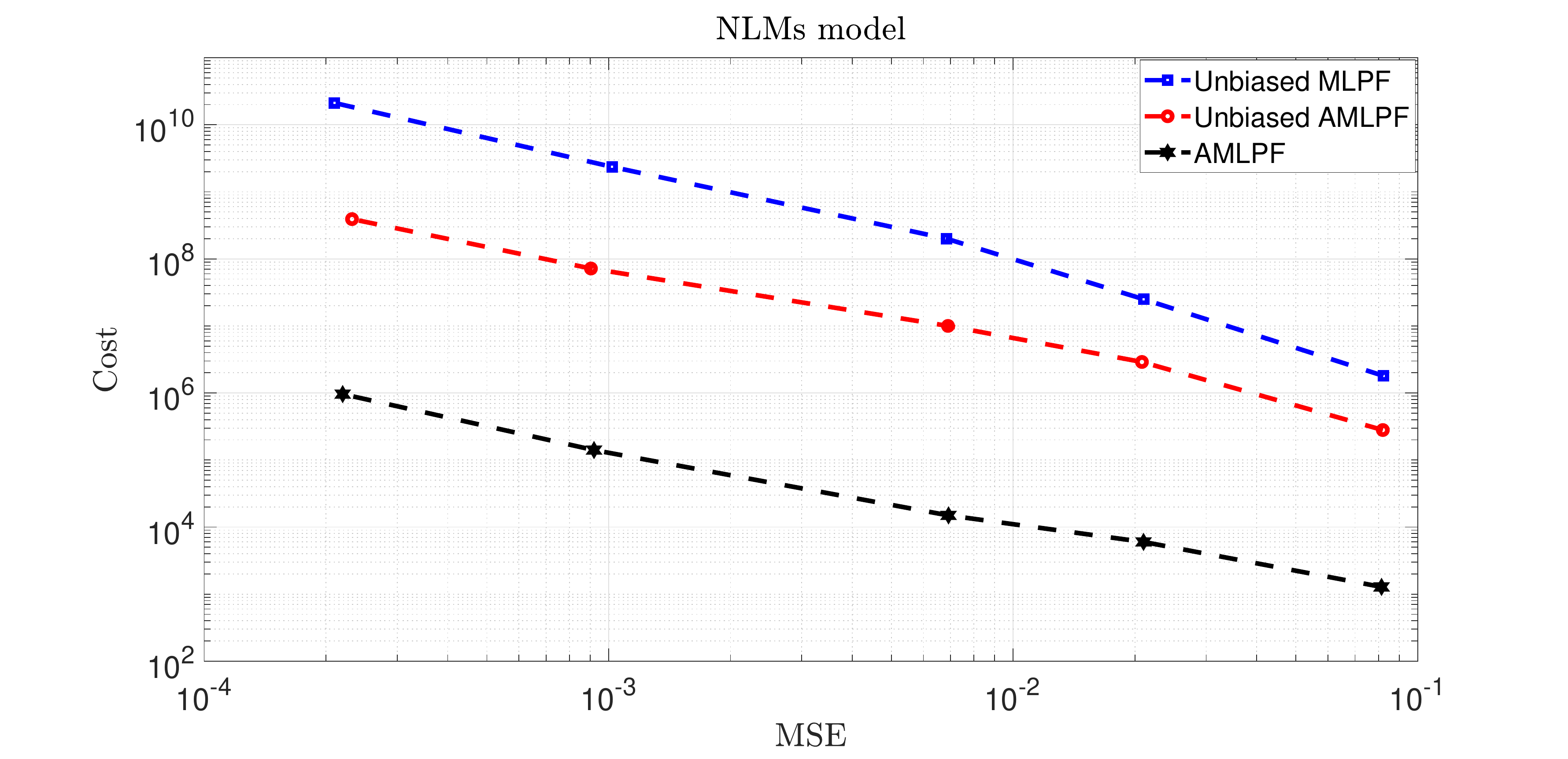}}
 \caption{Plots of cost rates as a function of the mean squared error.}
    \label{fig:MSEvsCost}
\end{figure}

The objective is to compare the costs of the three estimators for a given MSE.
Figure \ref{fig:MSEvsCost} presents our numerical results and we call the unbiased method
in \cite{ub_pf} as Unbiased MLPF (multilevel particle filter), our method as unbiased AMLPF (antithetic MLPF) and the method of \cite{anti_mlpf} as AMLPF. Each plot in the figure \ref{fig:MSEvsCost} is on the $\log_{10}$ scale.
Our results are run on a single core, so we do not take advantage of parallelization for the unbiased methods.
The simulation results show that as we reduce the MSE, the difference in the cost between the methods is higher. 
As expected for a single core, the AMLPF performs the best, but this gap can be easily reduced with parallelization and was discussed at length in \cite{ub_pf}.
The rates in Table \ref{tab:tab1}, associated to Figure \ref{fig:MSEvsCost}, confirm what was discussed in Section \ref{sec:math}. That is, for the unbiased MLPF we would expect cost rates around -1.25 (in fact the rates seem larger), for the unbiased AMLPF and AMLPF, we would expect around -1, which is what is seen.

\begin{table}[h!]
\begin{center}
\begin{tabular}{ c c c c } 
\hline \hline
Model  & Unbiased MLPF & Unbiased AMLPF & AMLPF   \\
\hline
\hline
\textbf{GBM} & -1.31 & -1.1  & -1.03  \\ 
 
\textbf{Clark-Cameron} & -1.42 & -1.16  & -1.09  \\
 
\textbf{NLMs} & -1.44 & -1.18  & -1.1  \\
 \hline
\end{tabular}
\caption{Estimated rates of MSE with respect to Cost. This is the log of the cost to log of MSE.}
\label{tab:tab1}
\end{center}
\end{table}

\subsubsection*{Acknowledgements}

All authors were supported by KAUST baseline funding.

\appendix

\section{Proofs}\label{app:intro}

To prove our results we use two major assumptions (A\ref{ass:diff1}) and (A\ref{ass:g}). These are given below.

\subsection{Some Notations}

Let $(\mathsf{V},\mathcal{V})$ be a measurable space.
For $\varphi:\mathsf{V}\rightarrow\mathbb{R}$ we write $\mathcal{B}_b(\mathsf{V})$ as the collection of bounded measurable functions. 
For $\varphi\in\mathcal{B}_b(\mathsf{V})$, we write the supremum norm $\|\varphi\|_{\infty}=\sup_{x\in\mathsf{V}}|\varphi(x)|$.
For a measure $\mu$ on $(\mathsf{V},\mathcal{V})$
and a function $\varphi\in\mathcal{B}_b(\mathsf{V})$, the notation $\mu(\varphi)=\int_{\mathsf{V}}\varphi(x)\mu(dx)$ is used. For $A\in\mathcal{V}$, the dirac measure is written as $\delta_A(dx)$.
If $K:\mathsf{V}\times\mathcal{V}\rightarrow[0,\infty)$ is a non-negative operator and $\mu$ is a measure, we use the notations
$
\mu K(dy) = \int_{\mathsf{V}}\mu(dx) K(x,dy)
$
and for $\varphi\in\mathcal{B}_b(\mathsf{V})$, 
$
K(\varphi)(x) = \int_{\mathsf{V}} \varphi(y) K(x,dy).
$
We denote (throughout) $C$ as a generic finite constant whose value may change upon each appearance and whose dependencies (on model and simulation parameters) are clear from the statements associated to them.

We write $\mathsf{X}_2=\mathbb{R}^{d\times d}$. $\mathbb{E}$ denotes the expectation w.r.t.~the law of our simulated algorithm.
The assumptions are as follows.
\begin{hypA}\label{ass:diff1}
\begin{itemize}
\item{For each $(i,j)\in\{1,\dots,d\}$, $\alpha_i\in\mathcal{B}_b(\mathsf{X})$, $\beta_{ij}\in\mathcal{B}_b(\mathsf{X})$.}
\item{$\alpha\in\mathcal{C}^2(\mathsf{X},\mathsf{X})$, $\beta\in\mathcal{C}^2(\mathsf{X},\mathsf{X}_2)$.} 
\item{$\beta(x)\beta(x)^{\top}$ is uniformly positive definite.}
\item{There exists a $C<+\infty$ such that for any $(x,i,j,k,m)\in\mathsf{X}\times\{1,\dots,d\}^{4}$:
$$
\max\left\{\left|\frac{\partial \alpha_i}{\partial x_m}(x)\right|,
\left|\frac{\partial \beta_{ij}}{\partial x_m}(x)\right|,
\left|\frac{\partial h_{ijk}}{\partial x_m}(x)\right|,
\left|\frac{\partial^2 \alpha_i}{\partial x_{k}\partial x_m}(x)\right|,
\left|\frac{\partial^2 \beta_{ij}}{\partial x_{k}\partial x_m}(x)\right|
\right\} \leq C.
$$}
\end{itemize}
\end{hypA}

\begin{hypA}\label{ass:g}
\begin{itemize}
\item{For each $k\in\mathbb{N}$, $g_k\in\mathcal{B}_b(\mathsf{X})\cap\mathcal{C}^2(\mathsf{X},\mathbb{R})$.}
\item{For each $k\in\mathbb{N}$ there exists a $0<C<+\infty$ such that for any $x\in\mathsf{X}$ $g_k(x)\geq C$.}
\item{For each $k\in\mathbb{N}$ there exists a $0<C<+\infty$ such that for any $(x,j,m)\in\mathsf{X}\times\{1,\dots,d\}^2$:
$$
\max\left\{\Big|\frac{\partial g_{k}}{\partial x_j}(x)\Big|,\Big|\frac{\partial^2 g_{k}}{\partial x_j\partial x_m}(x)\Big|\right\} \leq C.
$$
}
\end{itemize}
\end{hypA}

\subsection{Technical Results}

The following result is essentially  \cite[Proposition A.1]{ub_pf} and can be proved in the same manner.
\begin{lem}\label{lem:single}
Assume (A\ref{ass:diff1}-\ref{ass:g}). For any $k\in\mathbb{N}$ there exists a $C<+\infty$ such that for any
$(p,N_p,\underline{L},\varphi)\in\mathbb{N}_0\times\mathbb{N}\times\mathbb{N}_0\times\mathcal{B}_b(\mathsf{X})$:
$$
\mathbb{E}\left[\left(
\widehat{\pi}_k^{N_p,\underline{L}}(\varphi)-\pi_{k}^{\underline{L}}(\varphi)\right)^2\right]
\leq \frac{C\|\varphi\|_{\infty}^2}{N_p}\left(1+\frac{p^2}{N_p}\right)
$$
\end{lem}

Below we denote by $\eta_k^{s}$ as the discretized predictor at time $k\in\mathbb{N}$, at level $s\in\mathbb{N}_0$, that is, for any
$\varphi\in\mathcal{B}_b(\mathsf{X})$:
$$
\eta_k^s(\varphi) = \pi_{k-1}^l(P^l(\varphi))
$$
with the convention that $\pi_{-1}^s(dx)=\delta_{\{x_0\}}(dx)$.

\begin{lem}\label{lem:first_lem}
Assume (A\ref{ass:diff1}-\ref{ass:g}). For any $(k,\varphi)\in\mathbb{N}\times\mathcal{B}_b(\mathsf{X})\cap\mathcal{C}_b^2(\mathsf{X},\mathbb{R})$ there exists a $C<+\infty$ such that for any
$(p,N_p,l,\varepsilon)\in\mathbb{N}_0\times\mathbb{N}^2\times(0,\tfrac{1}{2})$:
$$
\mathbb{E}\left[\left(
[\tfrac{1}{2}\eta_k^{N_{0:p},l}+\tfrac{1}{2}\eta_k^{N_{0:p},l,a}-\eta_k^{N_{0:p},l-1}](\varphi) - [\eta_k^{l}-\eta_k^{l-1}](\varphi)
\right)^2\right]
\leq C\left(\frac{\Delta_l}{N_p}+\frac{p^2\Delta_l^{\tfrac{1}{2}-\varepsilon}}{N_p^2}\right).
$$
\end{lem}

\begin{proof}
This follows by combining the proof of \cite[Proposition A.1]{ub_pf} with \cite[Lemmata C.9 \& C.11]{anti_mlpf} along with some simple calculations which are omitted for brevity.
\end{proof}

\begin{rem}\label{rem:rem1}
One can prove, in a similar manner to Lemma \ref{lem:first_lem} the following results. For the first result, one needs 
\cite[Lemmata C.8 \& C.10]{anti_mlpf} and for the second \cite[Remarks C.4 \& C.5]{anti_mlpf}.
\begin{enumerate}
\item{Assume (A\ref{ass:diff1}-\ref{ass:g}). For any $(k,\varphi)\in\mathbb{N}\times\mathcal{B}_b(\mathsf{X})\cap\mathcal{C}_b^2(\mathsf{X},\mathbb{R})$ there exists a $C<+\infty$ such that for any
$(p,N_p,l,\varepsilon)\in\mathbb{N}_0\times\mathbb{N}^2\times(0,\tfrac{1}{2})$:
$$
\mathbb{E}\left[\left(
[\eta_k^{N_{0:p},l}-\eta_k^{N_{0:p},l-1}](\varphi) - [\eta_k^{l}-\eta_k^{l-1}](\varphi)
\right)^2\right]
\leq C\left(\frac{\Delta_l^{\tfrac{1}{2}}}{N_p}+\frac{p^2\Delta_l^{\tfrac{1}{2}-\varepsilon}}{N_p^2}\right).
$$
}
\item{Assume (A\ref{ass:diff1}-\ref{ass:g}). For any $(k,\varphi)\in\mathbb{N}\times\mathcal{B}_b(\mathsf{X})\cap\mathcal{C}_b^2(\mathsf{X},\mathbb{R})$ there exists a $C<+\infty$ such that for any
$(p,N_p,l,\varepsilon)\in\mathbb{N}_0\times\mathbb{N}^2\times(0,\tfrac{1}{2})$:
$$
\mathbb{E}\left[\left(
[\eta_k^{N_{0:p},l}-\eta_k^{N_{0:p},l,a}](\varphi)
\right)^2\right]
\leq C\left(\frac{\Delta_l^{\tfrac{1}{2}}}{N_p}+\frac{p^2\Delta_l^{\tfrac{1}{2}-\varepsilon}}{N_p^2}\right).
$$
}
\end{enumerate}
\end{rem}

\begin{lem}\label{lem:sec_lem}
Assume (A\ref{ass:diff1}-\ref{ass:g}). For any $(k,\varphi)\in\mathbb{N}\times\mathcal{B}_b(\mathsf{X})\cap\mathcal{C}_b^2(\mathsf{X},\mathbb{R})$ there exists a $C<+\infty$ such that for any
$(p,N_p,l,\varepsilon)\in\mathbb{N}_0\times\mathbb{N}^2\times(0,\tfrac{1}{2})$:
$$
\mathbb{E}\left[\left|
[\eta_k^{N_{0:p},l}-\eta_k^{N_{0:p},l-1}](\varphi) - [\eta_k^{l}-\eta_k^{l-1}](\varphi)
\right|^4\right]^{\tfrac{1}{4}}
\leq \frac{C 2^{\tfrac{p}{2}} \Delta_l^{\tfrac{1}{4}(\tfrac{1}{2}-\varepsilon)}}{N_p}.
$$
\end{lem}

\begin{proof}
Follows by using the definition of $\eta_k^{N_{0:p},s}$ with Minkowski's inequality $(p+1)$ times and then using \cite[Lemma C.8]{anti_mlpf} along with some simple calculations; the proof is omitted.
\end{proof}

\begin{rem}\label{rem:rem2}
Using a similar approach to the proof of Lemma \ref{lem:sec_lem} except using \cite[Remark C.4.]{anti_mlpf}
in place of \cite[Lemma C.8]{anti_mlpf} one can easily deduce the following result: Assume (A\ref{ass:diff1}-\ref{ass:g}). For any $(k,\varphi)\in\mathbb{N}\times\mathcal{B}_b(\mathsf{X})\cap\mathcal{C}_b^2(\mathsf{X},\mathbb{R})$ there exists a $C<+\infty$ such that for any
$(p,N_p,l,\varepsilon)\in\mathbb{N}_0\times\mathbb{N}^2\times(0,\tfrac{1}{2})$:
$$
\mathbb{E}\left[\left|
[\eta_k^{N_{0:p},l}-\eta_k^{N_{0:p},l,a}](\varphi)
\right|^4\right]^{\tfrac{1}{4}}
\leq \frac{C 2^{\tfrac{p}{2}} \Delta_l^{\tfrac{1}{4}(\tfrac{1}{2}-\varepsilon)}}{N_p}.
$$
\end{rem}

\begin{lem}\label{lem:coup}
Assume (A\ref{ass:diff1}-\ref{ass:g}). For any $(k,\varphi)\in\mathbb{N}\times\mathcal{B}_b(\mathsf{X})\cap\mathcal{C}_b^2(\mathsf{X},\mathbb{R})$ there exists a $C<+\infty$ such that for any
$(p,N_p,l,\varepsilon)\in\mathbb{N}_0\times\mathbb{N}^2\times(0,\tfrac{1}{2})$:
$$
\mathbb{E}\left[\left(
\widehat{[\pi_k^{l}-\pi_k^{l-1}]}^{N_p}(\varphi) - [\pi_k^{l}-\pi_k^{l-1}](\varphi)
\right)^2\right]
\leq C\left(\frac{\Delta_l}{N_p}+\frac{p^2\Delta_l^{\tfrac{1}{2}-\varepsilon}}{N_p^2}
+\frac{2^{2p}\Delta_l^{\tfrac{1}{2}-\varepsilon}}{N_p^4}
\right).
$$
\end{lem}

\begin{proof}
The result follows by combining Lemmata \ref{lem:first_lem},  \ref{lem:sec_lem} and Remarks \ref{rem:rem1} 
\ref{rem:rem2} with \cite[Lemma C.4.]{anti_mlpf} as we will now detail.  By using \cite[Lemma C.4.]{anti_mlpf}
along with the $C_2-$inequality 4 times we have the decomposition:
$$
\mathbb{E}\left[\left(
\widehat{[\pi_k^{l}-\pi_k^{l-1}]}^{N_p}(\varphi) - [\pi_k^{l}-\pi_k^{l-1}](\varphi)
\right)^2\right] \leq C\sum_{j=1}^4 T_j
$$
where:
\begin{eqnarray*}
T_1 & = & \mathbb{E}\Bigg[\Bigg(\frac{1}{\eta_{k}^{N_{0:p},l-1}(g_{k})}
[\tfrac{1}{2}\eta_{k}^{N_{0:p},l}+\tfrac{1}{2}\eta_{k}^{N_{0:p},l,a}-\eta_{k}^{N_{0:p},l-1}](g_{k}\varphi) -
\frac{1}{\eta_{k}^{l-1}(g_{k})}
[\tfrac{1}{2}\eta_{k}^{l}+
\tfrac{1}{2}\eta_{k}^{l}-\eta_{k}^{l-1}](g_{k}\varphi)
\Bigg)^2\Bigg]\\
T_2 & = &  \mathbb{E}\Bigg[\Bigg(\frac{1}{\eta_{k}^{N_{0:p},l}(g_{k})\eta_{k}^{N_{0:p},l-1}(g_{k})}\tfrac{1}{2}
\{[\eta_{k}^{N_{0:p},l}-\eta_{k-1}^{N_{0:p},l,a}](g_{k}\varphi)\}\{[\eta_{k}^{N_{0:p},l-1}-\eta_{k}^{N_{0:p},l}](g_{k})\}
\Bigg)^2\Bigg] \\
T_{3} & = & \mathbb{E}\Bigg[\Bigg(
\frac{\tfrac{1}{2}\eta_{k}^{N_{0:p},l,a}(g_{k}\varphi)}{\eta_{k}^{N_{0:p},l,a}(g_{k})\eta_{k}^{N_{0:p},l}(g_{k})\eta_{k}^{N_{0:p},l-1}(g_{k})}
\{[\eta_{k}^{N_{0:p},l,a}-\eta_{k}^{N_{0:p},l}](g_{k})\}\{[\eta_{k}^{N_{0:p},l-1}-\eta_{k}^{N_{0:p},l}](g_{k})\}
\Bigg)^2\Bigg] \\
T_{4} & = & \mathbb{E}\Bigg[\Bigg(
\frac{\eta_{k}^{N_{0:p},l,a}(g_{k}\varphi)}{\eta_{k}^{N_{0:p},l,a}(g_{k})\eta_{k}^{N_{0:p},l-1}(g_{k})}
[\tfrac{1}{2}\eta_{k}^{N_{0:p},l}+\tfrac{1}{2}\eta_{k}^{N_{0:p},l,a}-\eta_{k}^{N_{0:p},l-1}](g_{k}) - 
\frac{\eta_{k}^{l}(g_{k}\varphi)}{\eta_{k}^{l}(g_{k})\eta_{k}^{l-1}(g_{k})}\times \\ & &
[\tfrac{1}{2}\eta_{k}^{l}+\tfrac{1}{2}\eta_{k}^{l}-\eta_{k}^{l-1}](g_{k})
\Bigg)^2\Bigg].
\end{eqnarray*}
$T_1$ and $T_4$ can be treated in a similar manner, so we only consider $T_1$; this is the same for $T_2$ and
$T_3$ hence we only deal with $T_2$. Therefore we bound only $T_1$, $T_2$ and conclude the proof from there.

For $T_1$ we have that $T_1\leq C(T_5 + T_6)$ where
\begin{eqnarray*}
T_5 & = & \mathbb{E}\Bigg[\Bigg(\frac{1}{\eta_{k}^{N_{0:p},l-1}(g_{k})}\left(
[\tfrac{1}{2}\eta_{k}^{N_{0:p},l}+\tfrac{1}{2}\eta_{k}^{N_{0:p},l,a}-\eta_{k}^{N_{0:p},l-1}](g_{k}\varphi) -
[\tfrac{1}{2}\eta_{k}^{l}+
\tfrac{1}{2}\eta_{k}^{l}-\eta_{k}^{l-1}](g_{k}\varphi)
\right)\Bigg)^2\Bigg]\\
T_6 & = & 
 \mathbb{E}\Bigg[\Bigg(
\left(\frac{1}{\eta_{k}^{N_{0:p},l-1}(g_{k})}-\frac{1}{\eta_{k}^{l-1}(g_{k})}\right)
[\tfrac{1}{2}\eta_{k}^{l}+
\tfrac{1}{2}\eta_{k}^{l}-\eta_{k}^{l-1}](g_{k}\varphi)
\Bigg)^2\Bigg].
\end{eqnarray*}
$T_5$ can be controlled using the lower-bound on $g_k$ and Lemma \ref{lem:first_lem}. $T_6$ can be bounded
by using the lower-bound on $g_k$, \cite[Proposition A.1.]{ub_pf} and \cite[Lemma D.2.]{mlpf}. Putting these two results together gives
$$
T_1 \leq C\left(\frac{\Delta_l^{\tfrac{1}{2}}}{N_p}+\frac{p^2\Delta_l^{\tfrac{1}{2}-\varepsilon}}{N_p^2}\right).
$$
For $T_2$ we have that $T_1\leq C(T_7 + T_8)$ where
\begin{eqnarray*}
T_7 & = &  \mathbb{E}\Bigg[\Bigg(\frac{1}{\eta_{k}^{N_{0:p},l}(g_{k})\eta_{k}^{N_{0:p},l-1}(g_{k})}\tfrac{1}{2}
\{[\eta_{k}^{N_{0:p},l}-\eta_{k-1}^{N_{0:p},l,a}](g_{k}\varphi)\}\{[\eta_{k}^{N_{0:p},l-1}-\eta_{k}^{N_{0:p},l}](g_{k})
-\\ & &[\eta_{k}^{l-1}-\eta_{k}^{l}](g_{k})
\}
\Bigg)^2\Bigg]\\
T_8 & = &  \mathbb{E}\Bigg[\Bigg(\frac{1}{\eta_{k}^{N_{0:p},l}(g_{k})\eta_{k}^{N_{0:p},l-1}(g_{k})}\tfrac{1}{2}
\{[\eta_{k}^{N_{0:p},l}-\eta_{k-1}^{N_{0:p},l,a}](g_{k}\varphi)\}\{[\eta_{k}^{l-1}-\eta_{k}^{l}](g_{k})\}
\Bigg)^2\Bigg]
\end{eqnarray*}
For $T_7$ we can use the lower-bound on $g_k$, Cauchy-Schwarz,  Lemma \ref{lem:sec_lem}  and Remark
\ref{rem:rem2}. For $T_8$, we can use the lower-bound on $g_k$, Remark
\ref{rem:rem1} and \cite[Lemma D.2.]{mlpf}. Therefore, one can deduce that
$$
T_2 \leq C\left(\frac{\Delta_l}{N_p}+\frac{p^2\Delta_l^{\tfrac{1}{2}-\varepsilon}}{N_p^2}
+\frac{2^{2p}\Delta_l^{\tfrac{1}{2}-\varepsilon}}{N_p^4}
\right)
$$
and from here one can conclude.
\end{proof}

\end{document}